\documentclass[12pt]{article}

\usepackage{epsfig,rotate}
\usepackage{graphics}

\renewcommand{\phi}{\varphi}

\newcommand{\bnabla}{\mbox{\boldmath $\boldmath{\nabla}$\unboldmath }}

\begin{document}

\title{Kapitza resistance at the liquid solid interface\footnote{Paper submitted to the special issue
of Molecular Physics on the occasion of Dominique Levesque's 65th
birthday, edited by J-P. Hansen and R. Lynden-Bell} }

\author{Jean-Louis Barrat\footnote{author for correspondance; barrat@dpm.univ-lyon1.fr} and Fran\c{c}ois
Chiaruttini\\
D\'epartement de Physique des Mat\'eriaux\\ Universit\'e Claude
Bernard-Lyon I and CNRS\\ F-69622 Villeurbanne, France}


\maketitle

\begin{abstract}Using equilibrium and nonequilibrium
molecular dynamics simulations, we determine the Kapitza
resistance (or thermal contact resistance) at a model liquid solid
interface. The Kapitza resistance (or the associated Kapitza
length) can reach appreciable values when the liquid does not wet
the solid. The analogy with the hydrodynamic slip length is
discussed.
\end{abstract}

\section{Introduction: bulk and interfacial transport coefficients}

One of the important issues of nonequilibrium statistical physics
is the determination of transport coefficients, which relate the
flux of conserved quantities to the gradient of the corresponding
affinities. One of the simplest examples is the thermal
conductivity defined  macroscopically (in the absence of
concentration or velocity gradients) by
\begin{equation}
\bf{j_e} = - \lambda \bnabla{T} = L_e  \bnabla\frac{1}{T}.
\label{Fourier}
\end{equation}
Here $L_e$ is the Onsager coefficient associated with energy
transport, $1/T= \partial S/\partial E$ is the affinity relative
to energy.

Molecular dynamics, using either equilibrium or nonequilibrium
methods, has proven a powerful tool for determining such
coefficients, and, in fact, the only practical one when dealing
with dense liquids. Equilibrium methods for systems with
continuous potentials, pioneered by  Levesque and coworkers
\cite{VL}, are based on the use of Green-Kubo formulae
\cite{Hansen}. For the thermal conductivity, the formula reads:
\begin{equation}
\lambda= \int_0^\infty \langle \hat{j}_e(t) \hat{j}_e(0) \rangle
dt
 \label{Kubo1}
\end{equation}
where $\hat{j}_e$ is a microscopic energy current that can be
computed directly from the particle velocities and positions. The
brackets refer here to an equilibrium average on the microscopic
fluctuations {\it at equilibrium}. Green-Kubo calculations have
the advantage that an equilibrium simulation can be used to
determine simultaneously several transport coefficients. On the
other hand, rather long simulations are required to obtain
converged results for the time correlation function, and the
determination of the 'plateau' of the Green-Kubo integral (as a
function of the upper integration limit) is often difficult, since
at long times the noise in the correlation function tends to
produce a large numerical uncertainty.

A useful alternative for the determination of transport
coefficients is provided by nonequilibrium molecular dynamics
(NEMD) methods \cite{Hoover,Evans}. In such methods, the system is
driven into an equilibrium state using a modified MD algorithm,
and the currents and gradients are measured directly in this
state. Formally, these methods can be shown to be equivalent to
the Green-Kubo equilibrium method as long as one remains in the
linear response regime.

\begin{figure}
\center
\includegraphics[width=8cm,height=6.cm]{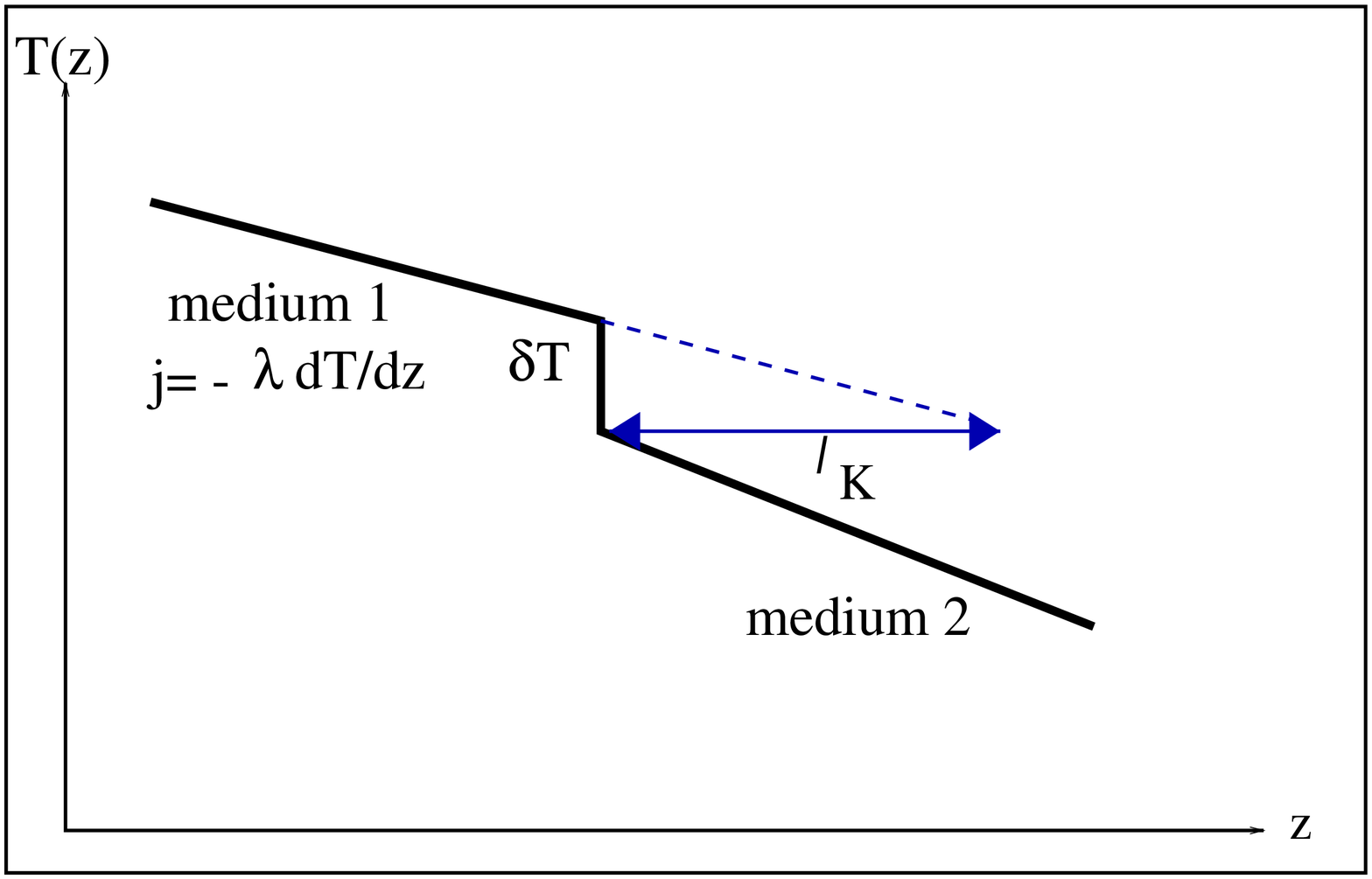}
\hspace{.3cm}
\includegraphics[width=8cm,height=6.cm]{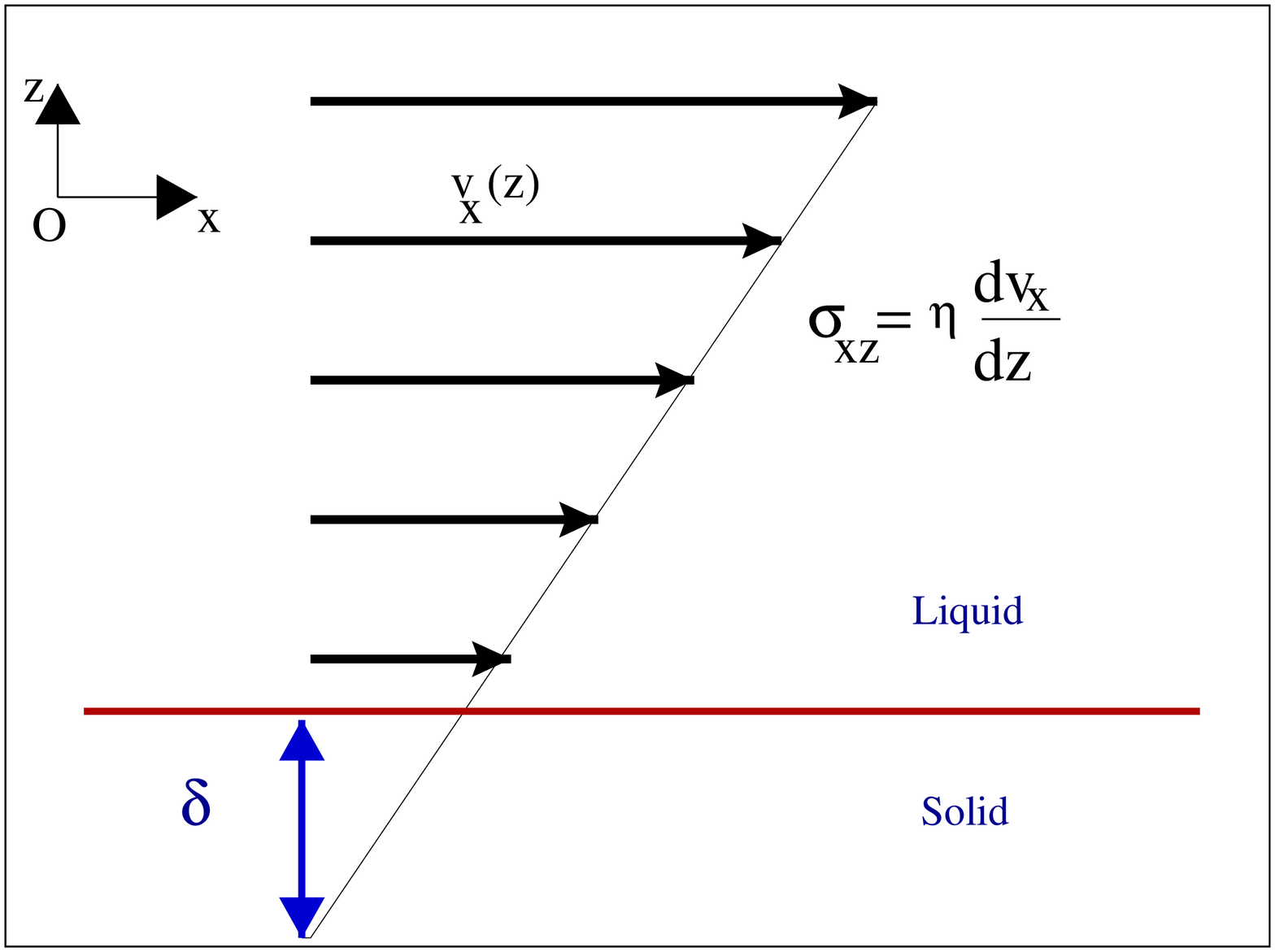}
  \caption{Schematic illustration of the Kapitza length (left) and
  of the slip length (right) at a liquid solid interface.}\label{fig1}
\end{figure}

In both types of approach, one usually is interested in {\it bulk}
transport coefficients, which are properties of the infinite,
homogeneous system. Much less attention has been devoted to
transport processes across interfaces between two bulk phases. At
the macroscopic scales, the flux of a conserved quantity must be
conserved across the interface. Continuum theories, however,
usually assume continuity not only of the fluxes, but also of the
affinities. The latter assumption, however, has no theoretical
justification, and is only known to work 'in practice' at the
macroscopic scale.  A more consistent description of transport
across an interface, in the spirit of usual hydrodynamic
descriptions,  allows for a jump in affinities between the two
phases, proportional to the flux of the conserved quantity. In the
case of energy transport, such a phenomenological description can
be translated mathematically as
\begin{equation}\label{rkdef}
  {\bf j}_E\cdot \bf{n}_{12} = \frac{1}{R_K}(T^{(1)} - T^{(2)}).
\end{equation}
To conform with usual notations, we have used the jump in
temperature rather than the jump in affinity $1/T$. Equation
\ref{rkdef} defines the {\it Kapitza resistance} $R_K$ (or thermal
contact resistance) of the interface, whose existence was pointed
out by Kapitza in the context of liquid Helium physics in
\cite{Kapitza}. A more transparent definition can be obtained by
defining the Kapitza length
\begin{equation}\label{defrk}
l_K= R_K \lambda
\end{equation}
where $\lambda$ is the thermal conductivity of one of the phases
(say phase $(1)$). $l_K$ is simply the thickness of material (1)
equivalent to the interface from a thermal point  of view (see
figure \ref{fig1}).
The Kapitza resistance has been mostly  studied in the case of
superfluid Helium in contact with a solid, or in the case of grain
boundaries between pure crystals
\cite{Kapitzahelium,kapitzacryst}. In both cases $\lambda$ is
large, so that the Kapitza length is large and measurable effects
can be expected. Size effects on the thermal conductivity of
composites are also an indirect way of assessing the thermal
properties of interfaces. In the case of normal liquids in contact
with a solid, however, it can be expected, on purely dimensional
grounds, that $l_K$ should be of the order of a few molecular
sizes, and therefore not measurable.

In this paper, we show that $l_K$ is indeed very small in the
-most usual- case of a wetting (contact angle smaller than
$\pi/2$) liquid-solid interface. In the nonwetting case, however,
the coupling between the solid and the liquid is strongly reduced,
and $l_K$ can be increased considerably - reaching up to 50
molecular sizes.

Before we discuss in more detail the determination of $l_K$ for a
model liquid-solid interface, it is interesting to point out an
analogy with another interfacial transport coefficient that has
attracted considerable attention in the recent years, namely the
'slip length'. The definition of the slip length completely
parallels that of the Kapitza length. In the bulk, the
phenomenological equation relates the momentum flux (stress) to
the velocity gradient (up to a factor of density, the velocity is
the affinity relative to momentum). The transport coefficient of
interest is the shear viscosity $\eta$, which in a simple, Couette
flow situation is defined by
\begin{equation}\label{defvisco}
  \sigma_{xz}= \eta \frac{\partial v_x}{\partial z}
\end{equation}
where $x$ is the flow direction and $z$ the direction of the
velocity gradient (see figure \ref{fig1}). At the interface, the
tangential velocity is not continuous in general. Since $v_x$ is
zero in a solid at rest, the equivalent of equation \ref{defrk}
reads
\begin{equation}\label{defkappa}
\sigma_{xz}(z=0) = \kappa v_x(z=0).
\end{equation}
Combining the two equations \ref{defvisco} and \ref{defkappa}, one
obtains a boundary condition for the velocity field
\begin{equation}\label{bcv}
  \frac{\partial v_x}{\partial z} = \frac{\kappa}{\eta} v_x(z=0)
  =\frac{1}{\delta}v_x(z=0).
\end{equation}
Equation \ref{bcv} defines the slip length $\delta$, which is the
distance over which the velocity field must me extrapolated inside
the solid to obtain a zero velocity (see figure \ref{fig1}). In
macroscopic hydrodynamics, $\delta$ is  usually assumed to vanish,
which amounts to a 'stick' boundary condition at the interface.
However, with the development of accurate surface force
measurement at the nanometer scale, it has become possible to
measure nonzero values of $\delta$. From a theoretical standpoint,
it has been possible to determine $\delta$ (or rather the
coefficient $\kappa$ in equation \ref{defkappa}) using a Kubo like
formula:
\begin{equation}\label{deltakubo}
\kappa= \frac{\eta}{ \delta} = \frac{1}{Sk_B T} \int_0^\infty
\langle F_{x}^{S\rightarrow L}(t) F_{x}^{S\rightarrow L}(0)
\rangle
\end{equation}
where $F_{x}^{S\rightarrow L}$ is the instantaneous force along
$x$ exerted by the liquid on the solid across the interface. It
has also been shown, both theoretically and experimentally, that
the wetting properties are an important ingredient in determining
$\delta$. Everything (lattice constants, densities) being
otherwise equal, changing the solid-liquid interaction (and
therefore the wettability) changes $\delta$ by orders of
magnitude. Large (50 nm or more) values of $\delta$ have been
measured for water on hydrophobic surfaces
\cite{churaev,crassous,cottin,granick}.

Our model of the liquid-solid interface will be identical to the
one used for our previous study of slip lengths. The interactions
between species (i) and (j)  are of the Lennard-Jones form
\begin{equation}\label{vij}
  v_{ij}(r) = 4\epsilon\left[
\left(\frac{\sigma}{r}\right)^{12} - c_{ij}
 \left(\frac{\sigma}{r}\right)^6\right]
\end{equation}
in which, for convenience, the strength of the attractive term is
modulated by a coefficient $c_{ij}$, keeping $\epsilon$ and
$\sigma$ identical (obviously the same effect could be achieved by
introducing different values of $\epsilon$ and $\sigma$; in this
form, emphasis is put on the attractive part which is essential in
controlling the wetting properties). For the liquid-liquid
interactions, we set $c_{11}=1.2$. For the FCC solid, $c_{22}=1$,
and the solid atoms are tethered to fixed lattice sites at a
constant density, $\rho_s \sigma^3=0.9$, using weak harmonic
springs. All interactions are cutoff  and shifted at
$2.5\sigma$. The interfacial properties of the (100 FCC solid with
the liquid have been studied in detail in reference \cite{BB}. By
varying the coefficient $c_{12}$, it is possible to vary the
contact angle $\theta$, up to $\theta=140^o$ for $c_{12}=0.5$.

In the next two  sections, we describe the methods used for
determining the Kapitza resistance of this model interface.
Results are presented in section \ref{results}.

\section{Nonequilibrium molecular dynamics simulations}

\begin{figure}
\center
\includegraphics[width=8cm]{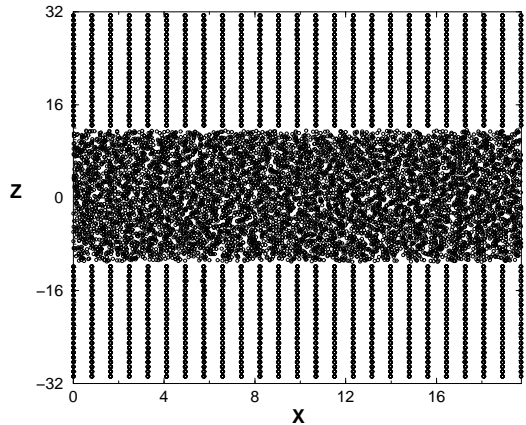}
  \caption{Transverse cut of a typical simulation cell. In order to avoid any
  possible problem with dissolution of solid atoms, those atoms are
  loosely bound to their equilibrium position through weak harmonic springs.}\label{fig2}
\end{figure}

The setup used for computing the Kapitza resistance is illustrated
in figure \ref{fig2}.
A slab of liquid  (typically of thickness between $30\sigma$ and
$40\sigma$), parallel to $xOy$ is sandwiched between two layers of
FCC solid of thickness $10-20\sigma$. The lateral dimensions of
the cell, depending on the simulation, have been varied between
$10$ and $20\sigma$. Periodic boundary conditions are applied in
the $x$ and $y$ directions. Liquid densities are typically of
order $\rho_l \sigma^3 =0.8-1$, and the reference temperature is
always $T=1\epsilon/k_B$. The normal pressure $P$ is obtained from
the average force exerted by the liquid on the solid; all the
simulations have been carried out at low values of the pressure,
$|P|<0.1\epsilon/\sigma^3$.

A temperature gradient is generated throughout the cell by
coupling the upper and lower atomic layers of the solid to two
thermostats  at different temperatures. The thermostats use simple
rescaling of the velocities for the group of atoms under
consideration. When the two temperatures are different, a net
energy flux $j_E$ in the $z$ direction results, which is computed
by averaging the kinetic energy added (or removed) by each
thermostat per unit time and surface. Temperature profiles are
obtained from the local kinetic energy density.

\begin{figure}
\center
\includegraphics[width=8cm]{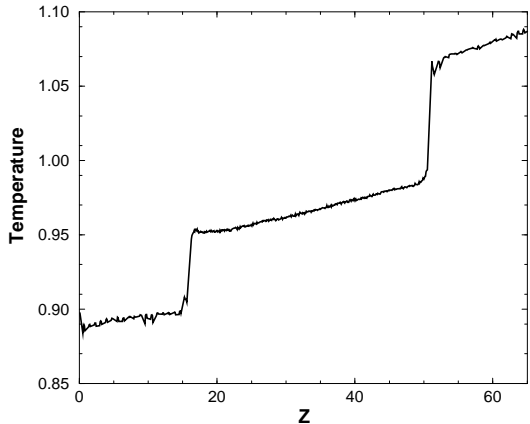}
  \caption{Temperature profile obtained for $c_{12}=0.5$}\label{fig3}
\end{figure}

A typical temperature profile obtained under these nonequilibrium
conditions  is shown in figure \ref{fig3}.
 In the 'bulk' liquid
and solid, the temperature varies essentially linearly with
position, in agreement with the predictions of Fourier's law. Note
that a slight curvature is perceptible in the liquid phase, which
reflects the dependence of $\lambda$ on temperature. From the
slope of the linear parts, the following values of the thermal
conductivity in the bulk phases are obtained (all results are
given in Lennard-Jones units, $\epsilon$, $\sigma$, and
$\tau=\sqrt{m\sigma^2/\epsilon}$);
\begin{equation}\label{vallambda}
  \lambda_l(T=1,P=0) =   7.8 \pm 0.5  \ \ \ ; \
   \lambda_s(T=1,\rho=0.9) =   8\pm 1
\end{equation}
The liquid has a density $\rho_s\sigma^3=0.88$ at $T=1$ and $P=0$.
The uncertainty on the results has two origins: first, the energy
current is the nonzero average value of a strongly fluctuating
quantity (i.e. the power extracted by the thermostats), and is
therefore difficult to determine accurately. Our simulations, with
a time step $h=0.01$ and typically $10^6$ steps, yield this
quantity with an accuracy of about 5\%. In the solid, the limited
thickness makes the determination of the temperature gradient
inaccurate, so that the uncertainty is somewhat larger than in the
liquid.

Another issue concerning the solid thermal conductivity is the
comparison of the phonon mean free path $\ell_{ph}$ with the
thickness of the solid slab. Using $\lambda_s \simeq \ell_{ph} u
c_v/3$ where $u$ is the sound velocity (obtained from the equation
of state)  and $c_v$ the heat capacity per unit volume, one finds
$\ell_{ph}\simeq 2.5\sigma$. Therefore the thickness of the solid
slabs is of order $3-4\ell_{ph}$,  which means that ballistic
effects in the transport of heat should be negligible. Indeed, we
confirmed by some simulations using thicker solid boundaries
($20\sigma$) that the results were not sensitively affected by
this parameter.

 From
the profile shown in figure \ref{fig3} the Kapitza resistance is
easily obtained by dividing the  temperature jump (computed using
the linear fits to the density profiles) by the energy flux. Note
that a slight dependence of $R_K$ on $T$ is visible in figure
\ref{fig3}. with a larger temperature jump on the 'warm' side. All
results discussed  in section \ref{results} will correspond to the
average values of $R_K$ between the 'warm' and 'cold' sides, and
therefore to  a nominal temperature $T=1$.

\section{Equilibrium determination of $R_K$}

A completely different route to the Kapitza resistance is provided
by a Green-Kubo formula similar to that used for the slip length,
equation \ref{deltakubo}. This formula, which to our knowledge was
first written by Puech et al \cite{Kapitzahelium}, reads
\begin{equation}\label{rkkubo}
\frac{1}{R_K}=\frac{1}{Sk_BT^2} \int_0^\infty dt \langle q(t) q(0)
\rangle
\end{equation}
Here $q$ is the energy flux across the interface of surface $S$,
which is macroscopically defined by the surface integral $q(t)=
\int\int {\bf j}_E(t)\cdot \mathbf{dS}$. Microscopically, $q(t)$
is easily computed from the work per unit time exerted by the
atoms belonging to the solid phase on those belonging to the
liquid phase, namely:
\begin{equation}\label{phit}
  q(t) = \sum_{i\in \mathrm{liquid}} \sum_{j \in \mathrm{solid}}
  \mathbf{F}_{ij}\cdot \mathbf{v}_{i}
\end{equation}

Equation \ref{rkkubo} can be justified through the following
heuristic  derivation. Let us consider a solid and a liquid
coupled together through an interface of surface $S$. The total
system (solid+liquid) is isolated, at  equilibrium. For
convenience, we will assume that the solid constitutes a
thermostat (infinite heat capacity) at fixed temperature $T_0$,
while the  heat capacity of the liquid part is denoted by $C_V$.
If $E_l(t)$ is the internal energy of the liquid phase at time
$t$, and if one assumes a Langevin evolution of this variable, one
may write
\begin{equation}\label{dedt}
\frac{dE_l}{dt}= q(t) = \frac{S}{R_K} (T_0-T_l(t)) + f(t)
\end{equation}
where $T_l$ refer to the instantaneous temperature of the liquid.
As usual in a Langevin description, we have separated the
evolution of the slow variable (the energy) into a systematic
contribution which would correspond to a macroscopic evolution and
a random, $\delta$ correlated part $f(t)$. Writing $\langle f(t)
f(t^\prime) \rangle = \gamma \delta(t-t^\prime)$,
$X=E_l(t)-\langle E_l \rangle= C_V (T_l(t)-T_0)$ one may rewrite
\ref{dedt} in the standard Langevin form
\begin{equation}\label{dxdt}
\frac{dX}{dt} = -\alpha X(t) + f(t)
\end{equation}
 with $\alpha= \frac{S}{R_K C_V}$. Equation \ref{dxdt}
 is a standard Langevin equation for the variable $X$.
The usual relations follow for the correlation functions
\begin{equation}\label{cx}
  \langle X(t) X(0) \rangle = \frac{\gamma}{2\alpha} \exp(-\alpha |t|)
\end{equation}
\begin{equation}
  \langle q(t) q(0) \rangle = -\frac{d^2}{dt^2}  \langle X(t) X(0)
  \rangle =  \gamma \delta(t) -\frac{\alpha \gamma}{2} \exp(-\alpha |t|)
\end{equation}
In the thermodynamic limit ($C_V\rightarrow \infty$) the second
term vanishes, so that one may write:
\begin{equation}\label{alpha}
\gamma = 2\int_0^\infty  \langle q(t) q(0) \rangle dt
\end{equation}
Using the relations $\langle X^2  \rangle =
\frac{\gamma}{2\alpha}$ which follows from equation \ref{cx} at
$t=0$, and the equilibrium distribution for $X$ which yields the
standard fluctuation formula $\langle X^2 \rangle = \langle \delta
E^2 \rangle = k_BT^2 C_V$, the Kubo relation \ref{rkkubo} for
$R_K$ is obtained  immediately from \ref{alpha}.

\begin{figure}
\center
\includegraphics[width=8cm]{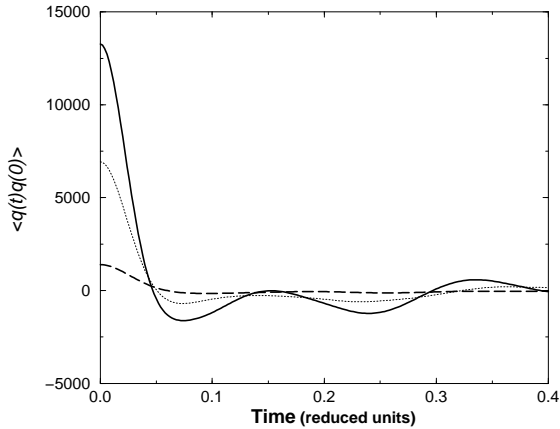}
  \caption{Time correlation functions for $q(t)$ for
  $c_{12}=0.5$, $c_{12}=0.8$, $c_{12}=1$ (the surface of the cell is $S=390\sigma^2$).}\label{fig4}
\end{figure}

\begin{figure}
\center
\includegraphics[width=8cm]{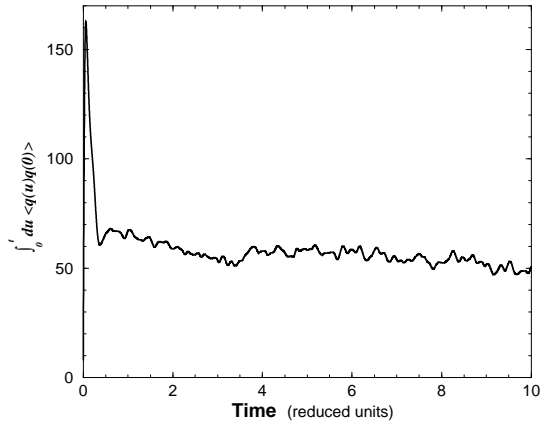}
  \caption{Running integral of the time correlation
  function $\langle q(t)q(0) \rangle$ obtained for $c_{12}=0.5$
  and $P=0.008$. A value of $R_K=6.9$ results, to be compared with
  the value $R_K=6.7$ obtained from a nonequilibrium simulation. }\label{fig5}
\end{figure}

As usual \cite{chaikin}, this line of reasoning can be extended to
the 'generalized Langevin' case where the coefficient $\alpha$ is
replaced by a memory kernel, and $f(t)$ is not a white noise. The
Kubo formula is not modified, and the macroscopic transport
coefficient being related to the time integral of the memory
Kernel. The Langevin approach just consists in assuming a
separation of time scales, so that the kernel can be replaced by a
$\delta$ function with the appropriate intensity. Note, however,
that the Kubo formula \ref{alpha} holds only {\em after} the
thermodynamic limit $C_V\rightarrow \infty$ has been taken. Hence,
for a finite system, the running integral $2\int_0^T \langle q(t)
q(0) \rangle dt$ will consist of two parts: a first part yielding
to a plateau value which contains the information on the Kapitza
resistance, and a long time decay to zero. Typical correlation
functions obtained in MD simulations (using the microscopic
definition \ref{phit}), and the corresponding time integrals, are
shown in figures \ref{fig4} and \ref{fig5}. The plateau of the
time integral can be used for a determination of the Kapitza
resistance.

\section{Dependance of $R_K$ on  wetting properties}
\label{results}

\begin{figure}
\center
\includegraphics[width=8cm]{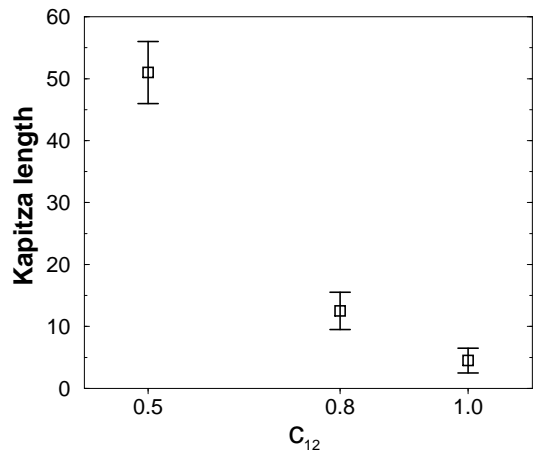}
  \caption{Kapitza length as a function of the interaction
  parameter $c_{12}$. All simulations are performed in the range
  of normal pressures $0.2>P>-0.2$, where we have checked that the
  pressure dependence of $l_K$ is weak. The error bars reflect the
  combination of several independent simulations (equilibrium and nonequilibrium)
  for determining each point. The heat conductivity of the liquid
  has been used to define the $l_K$ from the
  Kapitza resistance.}\label{fig6}
\end{figure}

It has been shown previously (see \cite{BB}) that the slip length
at the liquid solid interface was dependent on the wetting
properties. The ability of the solid to transfer momentum to the
liquid is weaker when the liquid does not wet the solid, since in
this case the liquid density is depleted in the vicinity of the
solid wall. Obviously the same kind of effect could be expected in
the Kapitza resistance, with a smaller rate of energy transfer
when the liquid-solid interaction is weak. In order to check this
expectation, we have studied, using both the nonequilibrium and
the equilibrium approach, the Kapitza resistance of our model
interface as a function of the interaction coefficient $c_{12}$,
which also determines the contact angle. All simulations are
performed at a pressure close to $P=0$. The results are shown in
figure \ref{fig6}, where it can be seen that the Kapitza length is
indeed a decreasing function of $c_{12}$, with values close to a
molecular size in the wetting case ($c_{12}=1$), and values up to
about 50 molecular sizes in the nonwetting case ($c_{12}=0.5$). As
a rule of thumb, $l_K$ appears to be of the same order of
magnitude as the slip length obtained in reference \cite{BB}.

\section{Summary}

In this paper, we have discussed the principles underlying the
definition of interfacial transport coefficients, and shown, in
the case of the Kapitza resistance, how such coefficients can be
extracted from equilibrium or nonequilibrium simulations. Our
results indicate that relatively large values of the Kapitza
length can be obtained using standard (as opposed to superfluid)
liquids, provided the liquid is not wetting the solid surface.
Experimentally, it seems very difficult to measure directly this
transport coefficient, as has been done for the slip length.
However, $R_K$ could probably be investigated by monitoring the
heat conductivity of a porous matrix filled with liquid.
Interfacial properties can then be tailored by treating the
internal surface of the pores (e.g. grafting of short hydrocarbon
chains to make the surface hydrophobic \cite{Benoit}).

{\bf Acknowledgments}: the parallel MD code 'LAMMPS2001' developed
by S. Plimpton (http://www.cs.sandria.gov/~sjplimp/lammps.html) was used throughout this work. All
calculations were performed at the PSMN at ENS-Lyon and at the
CDCSP of University Lyon I. Useful discussions with C.
Cottin-Bizonne, P. Chantrenne and L. Bocquet are acknowledged.



\begin{thebibliography}{199}

\bibitem{VL} D. Levesque, L. Verlet, J. K\"urkijarvi
Phys. Rev. A {\bf 7}, 1690 (1973).

\bibitem{Hansen} J.-P. Hansen and I.R. McDonald,
{\em Theory of simple liquids, 2nd edition} (Academic Press,
1986).

\bibitem{Hoover} W. G. Hoover, Ann. Rev. Phys. Chem. \textbf{34}, 103
(1983).


\bibitem{Evans} \textit{Statistical Mechanics of NonEquilibrium Liquids.} D.J. Evans, G.P.
Morriss. (Academic Press, London 1990).

\bibitem{Kapitza} P. L.~Kapitza,
J. Phys. USSR {\bf 4}, 181 (1941).

\bibitem{Kapitzahelium} L. Puech, G. Bonfait and B. Castaing,
Journal of Low Temperature Physics {\bf 62}, 312 (1986).

\bibitem{kapitzacryst}
R.J.~Stoner and H.J.~Maris, Physical Review {\bf B48}, 16373
(1993). A. Maiti, G.D. Mahan, S.T. Pantalides, Sol. State. Comm.
\textbf{102}, 517 (1997).

\bibitem{churaev} N.V.~Churaev, V.D.~Sobolev, and A.N.~Somov,
J. Coll. and Interf. Sci. {\em 97}, 574 (1984).


\bibitem{crassous} J. Baudry, A. Tonck, D. Mazuyer and E. Charlaix,
 Langmuir \textbf{17} 5232.

 \bibitem{cottin} C. Cottin-Bizonne, S. Jurine, J. Baudry, J. Crassous, E. Charlaix, F. Restagno
European Physical Journal E, in press.


\bibitem{granick} Y.~Zhu and S.~Granick,
Phys. Rev. Lett. {\em 87}, 96105-(1-4) (2001).

\bibitem{BB} J.-L.~Barrat and L.~Bocquet,
Faraday Discuss. {\bf 112}, 119 (1999); Phys. Rev. Lett.
\textbf{82}, 4671 (1999).

\bibitem{chaikin} P.M. Chaikin and T.C. Lubensky,
{\em Principles of condensed matter physics} (Cambridge University Press).

\bibitem{Benoit} T. Martin, B. Lefevre, L.Brunel {\it et al},
Chem. Commun. {\bf 1}, 24 (2002).


\end{thebibliography}
\end{document}